\documentclass{emulateapj}

\accepted{May 20, 2020}
\shorttitle{The Perpendicular Shock Acceleration of Cosmic Rays}
\shortauthors{Kamijima, Ohira and Yamazaki.}

\begin{document}

\title{Fast Particle Acceleration at Perpendicular Shocks with Uniform Upstream Magnetic Field and Strong Downstream Turbulence}

\author{Shoma F. Kamijima \altaffilmark{1}}
\email{s.kamijima@eps.s.u-tokyo.ac.jp}

\author{Yutaka Ohira\altaffilmark{1}}

\author{Ryo Yamazaki\altaffilmark{2,3}}

\altaffiltext{1}{Department of Earth and Planetary Science, The University of Tokyo, 
7-3-1 Hongo, Bunkyo-ku, Tokyo 113-0033, Japan}
\altaffiltext{2}{Department of Physics and Mathematics, Aoyama Gakuin University, 
5-10-1 Fuchinobe, Sagamihara 252-5258, Japan}
\altaffiltext{3}{Institute of Laser Engineering, Osaka University, 2-6 Yamadaoka,
Suita, Osaka 565-0871, Japan}




\begin{abstract}

The shock waves produce relativistic particles via the diffusive shock acceleration (DSA) mechanism.
Among various circumstances, fast acceleration has been expected for perpendicular shocks.
We investigate the acceleration time and the energy spectrum of particles accelerated at a perpendicular shock. 
In our model, the upstream perpendicular magnetic field has no fluctuation, and the downstream region is highly turbulent.
Then, the particle motion is the gyration in the upstream region and Bohm-like diffusion downstream.
Under this situation, we derive an analytical form of the acceleration time.
Using  test particle simulations,  the validity of our formula is verified.
In addition, the energy spectrum of particles  is the same as those predicted by standard DSA. 
Therefore, presently proposed mechanism simultaneously realizes the rapid acceleration 
and the canonical spectrum, $dN/dp\propto p^{-2}$, even if there is no upstream magnetic amplification.

\end{abstract}



\section{Introduction} \label{sec:intro}
The diffusive shock acceleration (DSA) is a probable mechanism of generating relativistic charged particles
at high-energy astrophysical phenomena like supernova remnants, which is the candidate of the origin of Galactic cosmic rays (CRs)\citep{axford77,krymsky77,bell78,blandford78}.
In DSA, particles are accelerated during the diffusive motion in which they are scattered by waves in the magnetized plasma.
The acceleration time depends on the angle between the shock normal and the background magnetic field
\citep{drury83},
and at first glance, it is smaller for perpendicular shocks than for parallel shocks \citep{jokipii87,giacalone99,shalchi09,shalchi10}.
In the former case, particles are confined just around the shock surface, and cannot escape out upstream.
In addition, they are more rapidly accelerated when the particle motion is less diffusive in a weaker turbulence,
although self-excited waves in the upstream region greatly reduce the acceleration rate \citep{zank04,zank06}.

There still remain problems on the acceleration at quasi- or exactly perpendicular shocks.
One of them is for the spectral slope of accelerated particles around the shock region.
Several authors numerically studied on this issue in various circumstances \citep{zank06,kong17,takamoto15}.
Generally speaking, in the case of weak turbulence to achieve rapid acceleration, the energy spectrum becomes steeper for the perpendicular shock acceleration 
than that predicted by the standard DSA theory, $dN/dp \propto p^{-2}$.
In \citet{takamoto15}, it is assumed that the magnetic fluctuation is weak both in the upstream and downstream regions, and that accelerated particles are isotropically scattered downstream.
Then, most particles are trapped by the background magnetic field and advected far downstream.
The probability for the particles to return upstream becomes smaller, leading to
softer energy spectrum than the standard DSA prediction.

X-ray observations of young SNRs have revealed  thin synchrotron filaments and/or time variability,
which implies the downstream magnetic field is amplified and turbulent 
\citep{vink03,bamba03,bamba05a,bamba05b,yamazaki04,uchiyama07}.
For example, gamma-ray and radio observations of Cas~A give us the lower limit of the downstream magnetic field strength of 120~$\rm \mu G$ \citep{ahnen17}.
Furthermore,   its hadronic gamma-ray spectrum shows that the maximum proton energy  is about 10~$\rm TeV$   \citep{ahnen17},
which leads the upstream magnetic field strength on the order of $\rm \mu G$
if the Bohm limit diffusion in both the upstream and the downstream regions is assumed.
The field amplification in the downstream region is also studied theoretically
\citep{giacalone07,ohira09b,inoue09,inoue12,caprioli13,ohira16a,ohira16b}.
Upstream density clumps hit the shock front,  generating downstream vorticities. 
Then, the field is twisted up and amplified.
On the other hand, upstream magnetic field configuration is uncertain.
In some cases as seen in Cas~A, the magnetic field strength and fluctuation are both as weak as interstellar medium.
If there are a number of accelerated particles in the upstream region, these accelerated particles interact with the background plasma, generating the turbulent magnetic field in the upstream region \citep{bell01,bell04,niemiec08,riquelme09,ohira09a}.
The particle injection rate can be less efficient at quasi-perpendicular shock for a fully ionized plasma \citep{gargate12,caprioli14}, but it can be efficient for a partially ionized plasma \citep{ohira12,ohira13,ohira16b}. 
In addition, if there is the pre-existing large-scale MHD turbulence in the upstream region, the particle injection can be effective even at quasi-perpendicular shocks \citep{giacalone05,zank06}.
In the present study, however, we consider the case that the upstream instabilities exciting turbulence is ineffective, resulting in the upstream fluctuation  much weaker than the  background magnetic field. Hence, as a first step, we assume that the upstream magnetic field is uniform.

Particles in the subluminal shock region can move along the magnetic field line and spread to the upstream region.
In this case, the upstream magnetic fluctuation becomes important. 
However, even though the upstream magnetic field is not uniform, it can be regarded as uniform as long as the shock becomes superluminal everywhere in the shock surface. 
The angle between the magnetic field and the shock normal direction is given by $\theta_{\rm Bn} \sim \tan^{-1}(B_0/\delta B)$, where $B_0$ and $\delta B$ are strengths of uniform perpendicular magnetic field and magnetic field fluctuations.  
The condition that the shock becomes superluminal is \citep{hoffmann50} 
\begin{eqnarray}
	\frac{u_{\rm sh}}{\cos \theta_{\rm Bn}} > v_{\parallel} \label{superluminal}~,
\end{eqnarray}
where $u_{\rm sh}$ and $v_{\parallel}$ are the shock velocity and the particle velocity parallel to the magnetic field. 
Therefore, the condition that we can ignore the effect of the magnetic field fluctuation in the perpendicular shock is given by 
\begin{eqnarray}
	\frac{\delta B}{B_0} < \frac{u_{\rm sh}}{c} \frac{ 1}{ \sqrt{1 - \left( \frac{u_{\rm sh}}{c} \right)^2} } ~, 
\end{eqnarray}
where $c$ is the speed of light. 
Since particles have velocities normal to the magnetic field and finite gyro radii, this condition can be relaxed.
Also note that the existence of shock ripples implies that $B_0$ has component parallel to (local) shock normal. This is, however, a subject beyond the current paper.
In the next paper, we will discuss in detail the case that there exists the magnetic fluctuation in the upstream region.

In this paper, we study the cosmic-ray acceleration at perpendicular shocks on the assumption that
 the upstream magnetic fluctuation is weak  and that the downstream region is highly turbulent.
 We show the acceleration time in the perpendicular shock and the energy spectrum under this assumption. 
This paper is organized as follows.
In Section \ref{sec:model}, we analytically estimate the acceleration time along with  our model.
In Section \ref{sec:test_particle_simulation}, we introduce our method of the test particle simulation.
The  results are presented in Section \ref{sec:result}.
Section \ref{sec:discussion} is devoted for summary and discussion.

\section{Model} 
\label{sec:model}
In the present study, it is assumed that the upstream magnetic field is uniform and has no fluctuation as an ideal case.
Then, the particles simply perform  gyro motion.
The downstream magnetic field is, on the other hand, highly turbulent, so that
downstream particles are transported by the isotropic Bohm diffusion in the downstream rest frame.
Under these conditions, we estimate the acceleration time for particles being energized at the perpendicular shock.
We use the notation ``1'' and  ``2''  representing  the upstream and the downstream regions, respectively.
The acceleration mechanism is DSA, which gives
the mean momentum gain per cycle $\Delta p/p =(4/3) (u_1 - u_2)/v$, where
$u_1$ and $u_2$ are the upstream and downstream flow velocities in the shock rest frame, respectively, and
$v$ is the particle velocity \citep{bell78}.
The acceleration time is calculated as $t_{\rm acc}=p/{\left(\Delta p/\Delta t\right)}$, where
$\Delta t=\Delta t_1 + \Delta t_2$ is a time for particles to make one cycle between upstream and downstream regions, and 
$\Delta t_1$ and $\Delta t_2$ are residence times in the upstream and downstream regions, respectively.
In this model, the upstream magnetic field is uniform and perpendicular to the shock normal direction. Then, the mean upstream residence time $\Delta t_1$ is determined by the gyro period:
\begin{eqnarray}
\Delta t_1 = \eta_{\rm g} \pi \Omega_{\rm g,1}^{-1} \label{eq:up_res}~~,
\end{eqnarray} 
where $\Omega_{\rm g,1} = qB_1/(\gamma mc)$ is the gyro angular frequency, and 
$m, q, \gamma,$ and $B_{\rm 1}$ are the particle mass, charge, Lorentz factor, and magnetic field strength in the upstream region.
A correction factor $\eta_{\rm g}$ is
on the order of unity and depends on the shock velocity $u_{\rm sh}$ and the particle velocity  $v$. 
The particle distribution and diffusion are assumed to be isotropic in the downstream region because of the strong magnetic turbulence. Then, the mean residence time in the downstream region $\Delta t_2$ is given by \citep{drury83}
\begin{eqnarray}
\Delta t_2 = \frac{4 D_2}{u_2 v} \label{eq:down_res}~~,
\end{eqnarray} 
where $D_2=\Omega_{\rm g,2}^{-1} v^2/3$ is the downstream diffusion coefficient  in the Bohm limit, and
$\Omega_{\rm g,2}$ is the gyro frequence in the downstream region.
The acceleration time is then calculated as
\begin{eqnarray}
t_{\rm acc,\perp} &=& \frac{3 \pi \eta_{\rm  g} r}{4 \left( r -1\right)} \left( \frac{u_{\rm sh}}{v} \right)^{-1} \Omega_{\rm g,1}^{-1}  \\ \nonumber
				&+& \frac{r^2}{r  -1} \left( \frac{B_2}{B_1} \right)^{-1} \left( \frac{u_{\rm sh}}{v} \right)^{-2} \Omega_{\rm g,1}^{-1}~~.
\label{eq:Tacc_our_model}
\end{eqnarray}
If the first term of right-hand-side dominates, we get $t_{\rm acc}\propto u_{\rm sh}{}^{-1}$,
while the standard DSA theory predicts $t_{\rm acc}\propto u_{\rm sh}{}^{-2}$
\citep{krymsky79,lagage81,drury83}.

The value of a correction factor $\eta_{\rm g}$ is almost unity for the following reason. 
The upstream residence time depends only on the gyrophase around the magnetic field line, $\phi$. 
Since the momentum distribution is isotropic in the shock downstream region, 
the number of particles in a range $\phi$ and $\phi + d \phi$ does not depend on $\phi$, that is, 
the distribution of the upstream residence time is almost uniform from 0 to $2\pi$ as long as $v\gg u_{\rm sh}$.
Then, the mean residence time becomes $\pi \Omega_{\rm g,1}^{-1}$.
Therefore, we simply set  $\eta_{\rm g}=1$ in this work.

In reality, the maximum value of the upstream residence time is smaller for larger shock velocity.
Since the shock front is moving in the upstream rest frame, 
the particle is caught up with the shock front while gyrating in the upstream region.
Particles are more rapidly caught up  by the shock front for higher shock velocity.
In addition, the particles with slower velocity component perpendicular to the shock surface than the shock velocity 
cannot return back to the upstream region.
This fact constrains the range of $\phi$ at which the particle cross the shock front to enter the upstream region \citep{achterberg01}. 
In the case of high shock velocity, relativistic effects become significant \citep{achterberg01,lemoine06,niemiec06}. 
In the present study,  we do not discuss the case of relativistically moving shocks.

Here, we compare the acceleration time of our model with that of DSA in the parallel shock, which is given by \citep{drury83}
\begin{equation}
t_{\rm acc,\parallel} = \frac{r}{r-1} \left\{ 1 + r\left( \frac{B_2}{B_1} \right)^{-1} \right\} \left( \frac{u_{\rm sh}}{v} \right)^{-2} \Omega_{\rm g,1}^{-1} ~~, 
\label{eq:tacc_para}
\end{equation}
where the Bohm diffusion is assumed both in the upstream and downstream regions. 
The difference between $t_{\rm acc,\perp}$ and $t_{\rm acc,\parallel}$ is the upstream residence time, $\Delta t_1$, which is given by 
$4D_1/u_{\rm sh}v$ for the parallel shock. 
Even in the case of Bohm diffusion, the upstream residence time in the parallel shock region is $(4/3 \pi)(v/u_{\rm sh})$ 
times longer than that of our model in the perpendicular shock for $u_{\rm sh}/v < 4/3 \pi \sim 0.4$. 
\begin{figure}[h]
	\centering
	\includegraphics[bb = 0 0 640 480, scale=0.39]{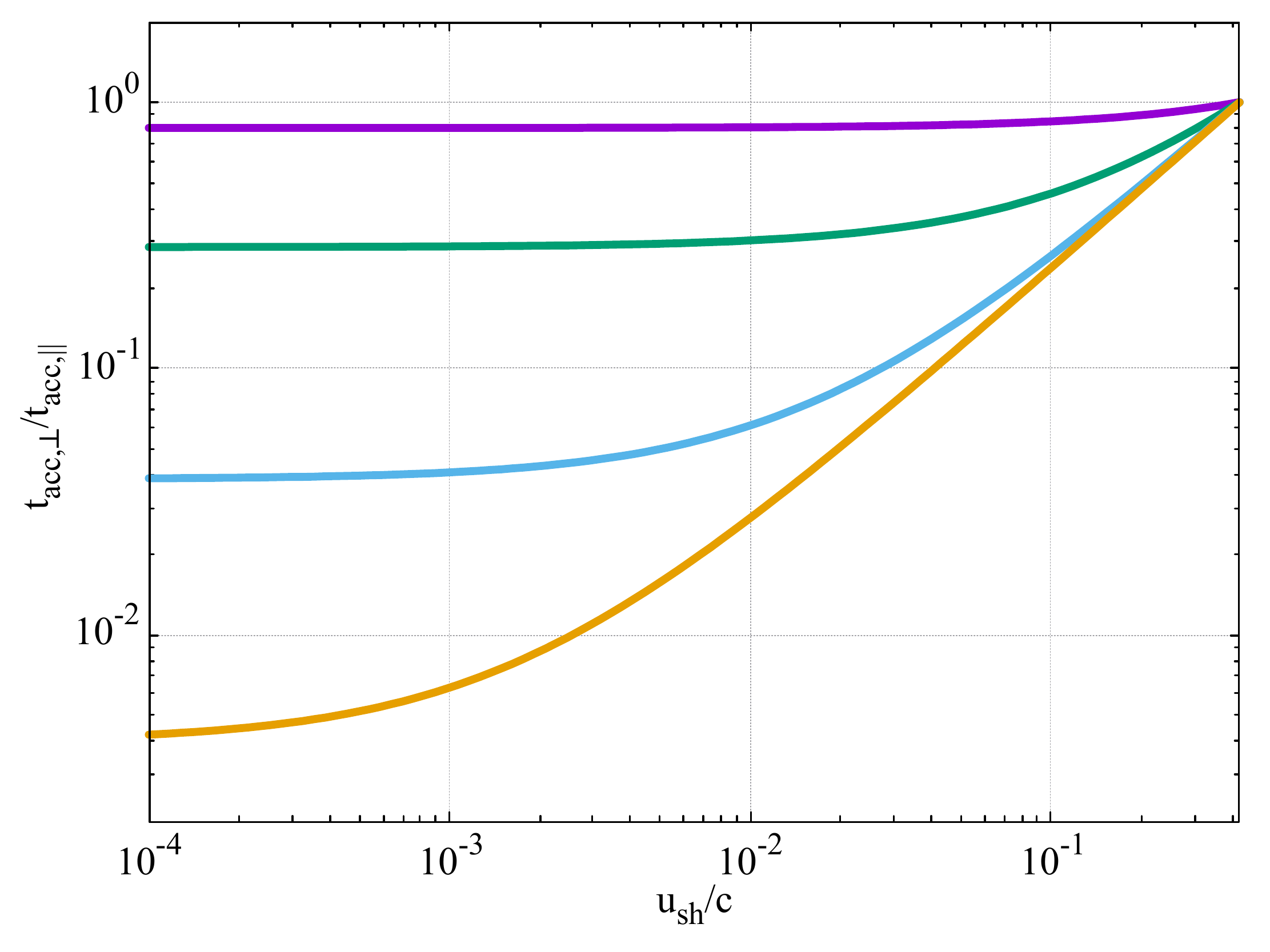}
	\caption{The ratio of the acceleration time of our model (equation (\ref{eq:Tacc_our_model})) and that  in a parallel shock (equation (\ref{eq:tacc_para})) as a function of the shock velocity $u_{\rm sh}/c$. 
We set the compression ratio $r=4$, the correction factor $\eta_{\rm g}=1$, and particle velocity $v=c$.
The purple, green, cyan, and orange lines show the ratio for the downstream field strength of $B_2/B_1=1,10,100,$ and $1000$, respectively.
\label{fig:tacc_ratio_wo_upmagamp}}
\end{figure}
Figure \ref{fig:tacc_ratio_wo_upmagamp} shows the ratio, $t_{\rm acc,\perp}/t_{\rm acc,\parallel}$, as a function of the shock velocity,
where we set the compression ratio $r=4$, the correction factor $\eta_{\rm g}=1$, and particle velocity $v=c$.
The purple, green, cyan, and orange lines show the ratio for the downstream field strength of $B_2/B_1=1,10,100,$ and $1000$, respectively.
The acceleration time of our model becomes shorter than that of the parallel shock case. 
The former is mainly determined by  
the downstream residence time if the shock velocity is sufficiently slow, 
so that the ratio, $t_{\rm acc,\perp}/t_{\rm acc,\parallel}$, closes to the constant value. 
On the other hand, for a sufficiently fast shock velocity, the upstream residence time in the parallel shock closes to the gyro period, 
so that the ratio, $t_{\rm acc,\perp}/t_{\rm acc,\parallel}$, approaches to unity. 
Note that the diffusion approximation is not valid for $u_{\rm sh}/c \gtrsim 0.4$ as long as the Bohm diffusion is assumed 
because the diffusion length becomes shorter than the gyroradius. 
The upstream magnetic field is expected to be amplified in the parallel shock \citep[e.g.][]{bell04}. 
For the Bohm diffusion, if the upstream magnetic field is amplified to more than $(4/3 \pi)(v/u_{\rm sh})$ times the initial magnetic field strength, 
particles can be accelerated in the parallel shock faster than our perpendicular model. 

In our model, in addition to the rapid acceleration, the same momentum spectrum as that of the standard DSA, $dN/dp \propto p^{-2}$, is expected because the downstream momentum distribution is assumed to be isotropic. 
To confirm the rapid acceleration and the canonical momentum spectrum, we perform test particle simulations in the next section.

\section{Test particle simulation} 
\label{sec:test_particle_simulation}
In this paper, we consider a plane shock wave, and the shock normal is along the $x$ axis. 
Upstream uniform magnetic field is taken as $\vec{B}_0=B_0 \hat{z}$. 
Different methods of particle transport are used in upstream and downstream regions, respectively.
In the upstream region, particle orbit is determined by solving the equation of motion of charged particles in the uniform magnetic field $\vec{B}_0$.
In the upstream rest frame, the equation of motion for particles with charge $q$  is given by
\begin{equation}
\frac{d \vec{p}}{dt} = q \left( \frac{\vec{v}}{c} \times  \vec{B}_0  \right)~~.
\end{equation}
The Bunemann-Boris method is used to solve this equation \citep{birdsall91}.
We use the Monte-Carlo method for the downstream particle transport.
Since the downstream magnetic field is amplified and turbulent, 
particles in the downstream region are isotropically scattered in the downstream rest frame.
The scattering time of accelerated particles is proportional to their momentum.
The particle splitting method is adopted to improve the statistics of the number of high energy particles.
Only at the initial time $t=0$, particles are homogeneously injected at the shock surface with the initial Lorentz factor $\gamma_0=15$.
The initial velocity distribution is isotropic in the velocity space.
The compression ratio $r$ is set to be 4 in the following.
Table \ref{table:parameters} shows parameters from Run~1 to Run~20.
\begin{table}[h]
	\caption{Parameters about the magnetic field amplification in the downstream region and the shock velocity \label{table:parameters}} 
	\begin{center}
	\normalsize
	\begin{tabular}{lrr} \hline \hline
		Run     & $B_2/B_1$ & $u_{\rm sh}/c$ \\ \hline 
		1   & 1 & 0.01 \\ \hline
		2   & 10     & 0.001 \\ 
		3   & 10     & 0.00316 \\
		4   & 10     & 0.01 \\ 
		5   & 10     & 0.0316 \\
		6   & 10     & 0.1 \\ 
		7   & 10     & 0.316 \\ \hline
		8   & 100   & 0.001 \\ 
		9   & 100   & 0.00316 \\
		10 & 100   & 0.01 \\ 
		11 & 100   & 0.0316 \\
		12 & 100   & 0.1 \\ 
		13 & 100   & 0.316 \\ \hline
		14 & 300 & 0.01 \\ \hline
		15 & 1000 & 0.001 \\
		16 & 1000 & 0.00316 \\
		17 & 1000 & 0.01 \\
		18 & 1000 & 0.0316 \\
		19 & 1000 & 0.1 \\ 
		20 & 1000 & 0.316 \\ \hline
	\end{tabular}
	\end{center}
\end{table}

\section{Simulation Results} 
\label{sec:result}
\subsection{Acceleration Time}
\begin{figure}[h]
	\centering
	\includegraphics[bb = 0 0 640 480, scale=0.39]{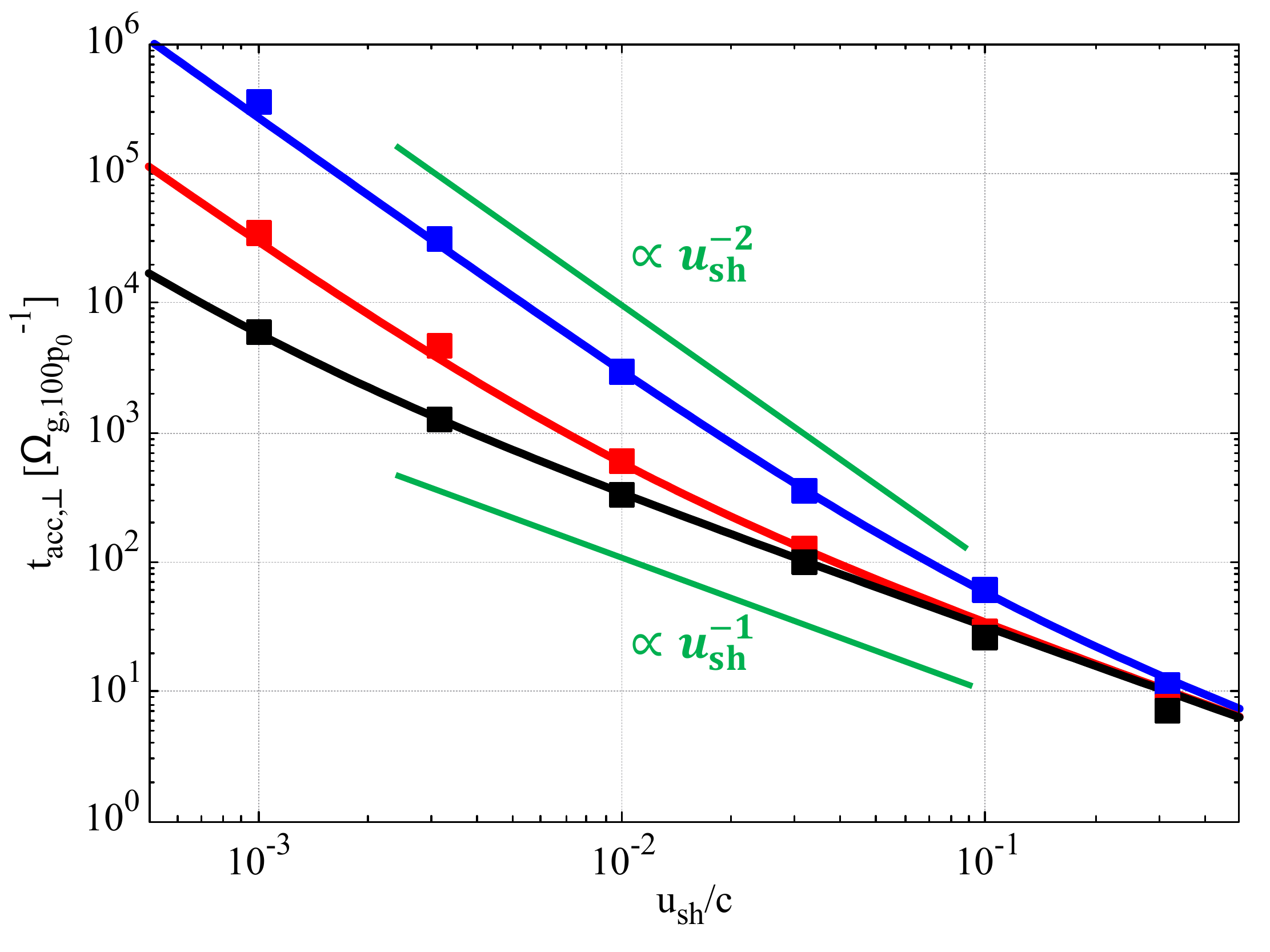}
	\caption{Acceleration time as a function of the shock velocity. The data points and lines show simulation results and the theoretical curves (equation (\ref{eq:Tacc_our_model})), respectively. The blue, red, and black colors show the simulation results for $B_2/B_1=10, 100, {\rm and}\ 1000$. The acceleration time and shock velocity are normalized by $\Omega_{\rm g,100p_0}^{-1}$ and $c$, respectively. \label{fig:tacc-ush_noturb}}
\end{figure}
We first investigate the acceleration time of particles with the momentum of $100\ p_0$. 
The acceleration time is defined by the elapsed time when the cutoff position of the momentum spectrum becomes 100 times as large as the initial momentum, $p_0$. 
The momentum spectra for Run~17 (red histogram) and Run~5 (blue histogram) at $t=5.35 \times 10^2 \Omega_{\rm g,100p_0}^{-1}$ are shown in Figure~\ref{fig:spectrum2_comparison}, where $\Omega_{\rm g,100p_0}$ is the gyro frequency of particles with the momentum of $100\ p_0$. 
In these cases, the cutoff momentum is estimated to be $145\ p_0$ and $103\ p_0$ for Run~17 (red histogram) and Run~5 (blue histogram), respectively. 
We obtain the momentum spectrum and estimate the cutoff momentum every at every time step, from which we identify the acceleration time when the cutoff momentum becomes $100\ p_0$.

Figure \ref{fig:tacc-ush_noturb} shows the acceleration time of particles with the momentum of $100\ p_0$ as a function of the shock velocity. 
The data points and lines show simulation results and the theoretical curves (equation (\ref{eq:Tacc_our_model})), respectively.
The blue, red, and black colors show the results for the downstream field strength $B_2/B_1=10, 100, {\rm and}\ 1000$, respectively.
As one can see, our theoretical curves of the acceleration time are in good agreement with simulation results. 
For a sufficiently high shock velocity, the downstream residence time becomes negligible compared with the upstream residence time, so that $t_{\rm acc}\propto u_{\rm sh}{}^{-1}$. 
On the other hand, for a sufficiently slow shock velocity, the downstream residence time becomes longer than that in the upstream region, 
so that $t_{\rm acc}\propto u_{\rm sh}{}^{-2}$. 
The transition from $t_{\rm acc}\propto u_{\rm sh}^{-1}$ to $t_{\rm acc}\propto u_{\rm sh}^{-2}$ occurs when the downstream residence time becomes equal to the upstream residence time.
Then, the condition $\Delta t_1=\Delta t_2$ gives the critical shock velocity which is given by   
\begin{equation}
\frac{u_{\rm sh,c}}{c} = \frac{3 \pi}{4 r} \left( \frac{B_2}{B_1} \right)^{-1}~~.
\label{eq:ushc}
\end{equation}
For $B_2/B_1=1, 10, 100~{\rm and}~1000$, we have $u_{\rm sh,c}/c=0.6$, 0.06, 0.006 and 0.0006, 
which are consistent with simulation results. 

\begin{figure}[h] 
	\centering
	\includegraphics[bb = 0 0 640 480, scale=0.39]{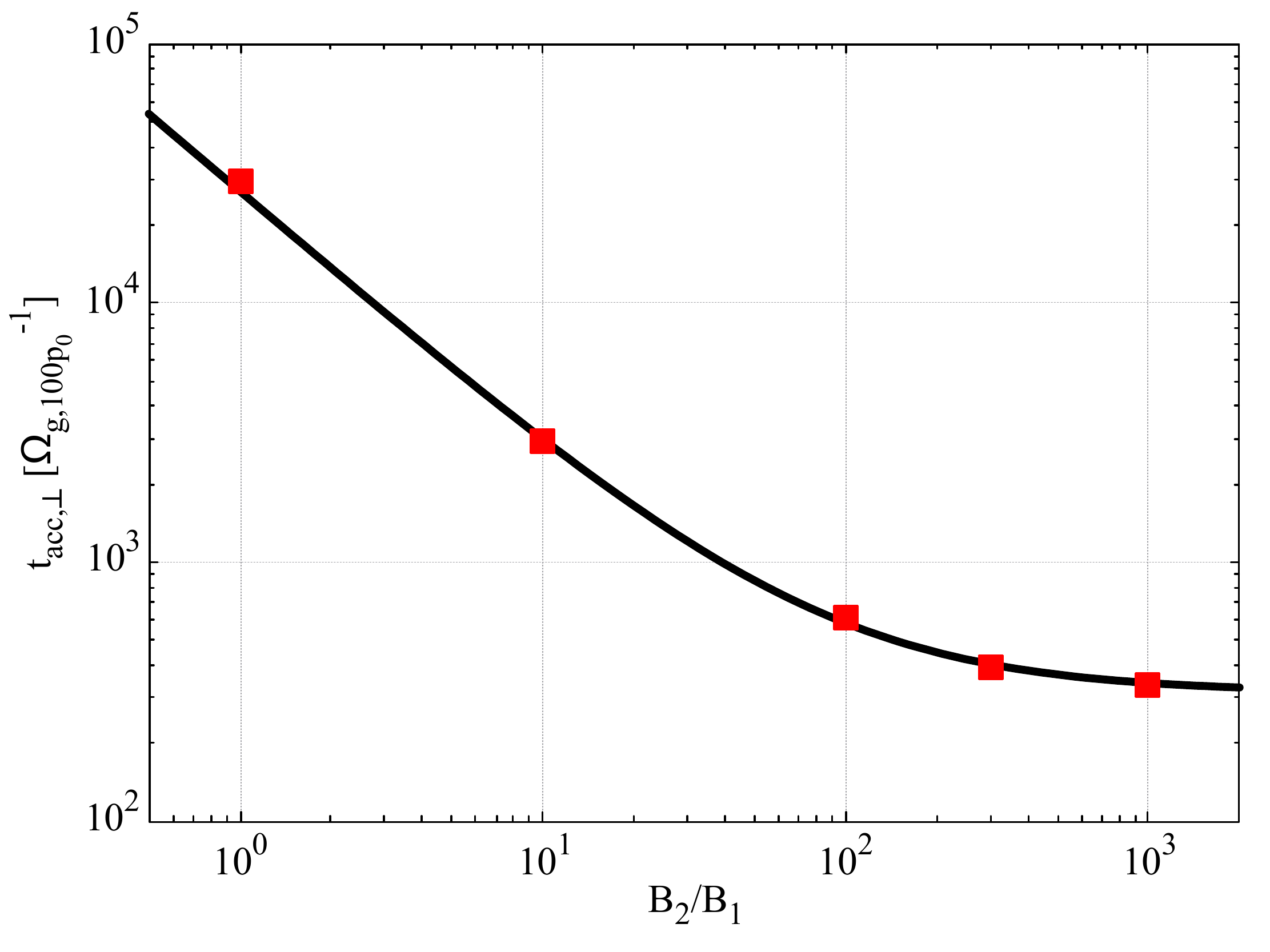}
	\caption{Acceleration time as a function of downstream magnetic field strength for $u_{\rm sh}/c=0.01$. 
The data points and line show simulation results and the theoretical curve (equation (\ref{eq:Tacc_our_model})), respectively. \label{fig:tacc-alpha_noturb}}
\end{figure}
Figure \ref{fig:tacc-alpha_noturb} shows the acceleration time of particles with the momentum of $100\ p_0$ as a function of $B_2/B_1$ in the case of $u_{\rm sh}/c=0.01$. 
Again our theoretical curve (solid curve) well explains the simulation results.
As the downstream magnetic field strength becomes large, the downstream residence time decreases, 
however, the upstream residence time remains unchanged.
Then, the acceleration time does not depend on the downstream magnetic field strength for  sufficiently large downstream magnetic field. 
Hence, the acceleration time is mainly determined by the upstream residence time when
\begin{equation}
\frac{B_2}{B_1} >  \frac{4 r}{3 \pi} \left( \frac{u_{\rm sh}}{c} \right)^{-1}~~.
\label{eq:b2c}
\end{equation}
For $u_{\rm sh}/c=0.01$, the above condition becomes $B_2/B_1>1.7\times10^2$,
which is consistent with our simulation result (see  Figure~\ref{fig:tacc-alpha_noturb}).

Therefore, these simulations confirm that as long as the downstream magnetic field is sufficiently amplified, particles are rapidly accelerated by the perpendicular shock compared with a case of the parallel shock without any upstream magnetic field amplification.

\subsection{Momentum Spectrum}
\begin{figure}[h]
	\centering	
	\includegraphics[bb = 0 0 640 480, scale=0.39]{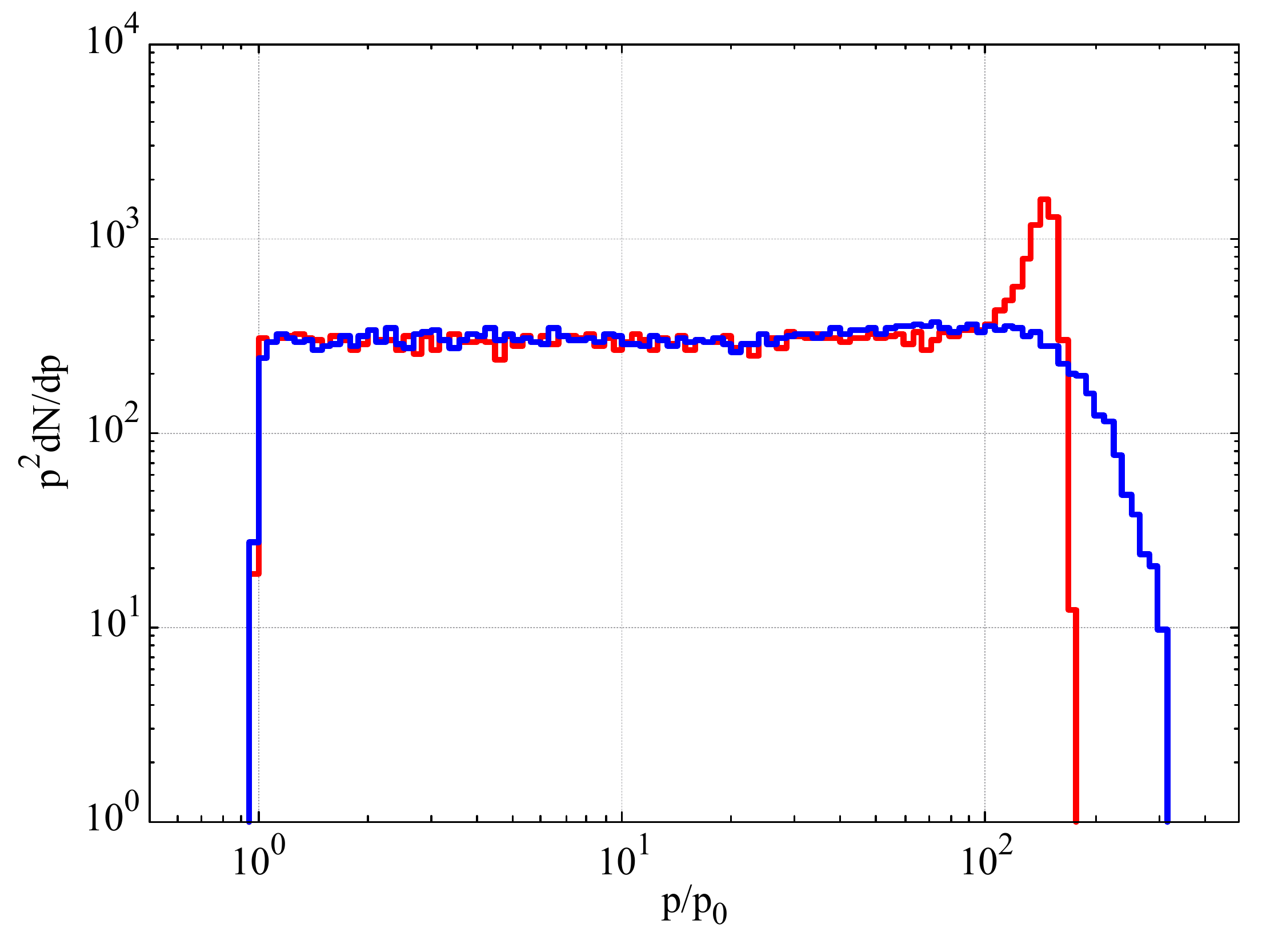}
	\caption{	Momentum spectra of all simulation particles for Runs 17 (red histogram) and 5 (blue histogram) at $t=5.35 \times 10^2 \Omega_{\rm g,100p_0}^{-1}$. 
		Parameters of Run~5 are $u_{\rm sh}/c = 0.0316$ and $B_2/B_1=10$, while
                 		$u_{\rm sh}/c = 0.01$ and $B_2/B_1=1000$ for Run~17.
		For Run 5 (17), the upstream residence time is shorter (longer) than the downstream residence time.
		\label{fig:spectrum2_comparison}}
\end{figure}
The momentum spectra derived by Run~17 (red histogram) and Run~5 (blue histogram) 
at $t=5.35 \times 10^2 \Omega_{\rm g,100p_0}^{-1}$ are shown in Figure~\ref{fig:spectrum2_comparison}.
The spectral indies in the power-law regime are consistent with  $-2.0$, which agrees with the prediction of 
standard DSA \citep{bell78}. 
In the other Runs,
 the spectral index is also consistent with $-2.0$.
Therefore, even in absence of upstream magnetic turbulence, 
the spectra of particles accelerated at the perpendicular shock are given by $dN/dp\propto p^{-2}$ 
as long as they are isotropically scattered downstream.

It was shown that the weaker the magnetic turbulence, the smaller acceleration time for the perpendicular shock, 
but at the same time, the steeper spectra of accelerated particles  than  the prediction of standard DSA \citep{takamoto15}. 
In \citet{takamoto15}, the magnetic field fluctuation is assumed to be weak in both the upstream and downstream regions. 
The weak fluctuation makes the return probability from the downstream region small for the perpendicular shock, so that the spectrum becomes steeper. 
On the other hand, in this study, we consider the case where the magnetic field fluctuation is weak in the upstream region but strong in the downstream region. 
Since the strong downstream turbulence  is responsible for isotropic particle distribution,
the weak turbulence ahead of the perpendicular shock 
with strongly turbulent downstream magnetic field can realize the rapid acceleration and the canonical spectrum, $dN/dp\propto p^{-2}$, simultaneously.

The simulation results show that the cutoff shape around the maximum momentum depends on whether the upstream residence time is shorter than the downstream residence time or not. 
For Run~17 (red histogram) where the upstream residence time is shorter than the downstream one ($\Delta t_1<\Delta t_2$), 
we identify a bump accompanied by  very sharp cutoff compared with the opposite case (Run~5: blue histogram). 
This is because the cutoff shape depends on the probability distribution of the residence time \citep{drury83}. 
The probability distribution of the upstream residence time is almost constant between 0 and $\pi \Omega_{\rm g,1}^{-1}$ in this study. 
On the other hand, the  probability distribution of the downstream residence time has a peak at a time much smaller than the mean residence time \citep{kato03}. 
Then, many particles can experience the back and forth motion at the shock front during the time scale shorter than the mean downstream residence time if the downstream residence time is longer than the upstream residence time.
As a result, there are many particles with the momentum larger than the cutoff scale which is decided by the mean residence time. 
The cutoff shape around the maximum energy in the supernova remnants could be more precisely determined by
future observations of synchrotron radiation and/or inverse Compton radiation from electrons with the maximum energy \citep{yamazaki14}. 
Thereby we could investigate whether particles are accelerated at the perpendicular shock or not. 

\section{Discussion} 
\label{sec:discussion}
\subsection{Broken power law spectrum}
If the magnetic field is amplified by the downstream turbulence, 
It takes a finite time to stretch the magnetic field line and to be turbulent in the downstream region. 
In addition, the amplified magnetic field eventually decays far downstream \citep{pohl05}. 
Thus, the magnetic field is not necessarily amplified and turbulent all over the shock downstream region. 
In this case, the magnetic field configuration would be similar to the condition of \citet{takamoto15}, leading to a  spectrum softer than that of the standard DSA.
Whether accelerated particles have the softer spectrum or not depends on where in the downstream region the accelerated particles are scattered during acceleration. 
The diffusion region depends on the momentum of the accelerated particles. 
Therefore, the momentum spectrum at the perpendicular shock does not necessarily have a single power law form. 
Broken power-law spectra observed in various astrophysical objects might be generated by the perpendicular shock. 

\subsection{Maximum energy limited by a finite age of supernova remnants}
We consider the maximum energy in the case where particles continue to be accelerated during the age of supernova remnants, $t_{\rm age}$. 
If the acceleration time is mainly determined by the upstream residence time, the maximum energy $E_{\rm max,age,1}$ is given by 
\begin{eqnarray}
	E_{\rm max,age,1} = \frac{4 (r-1) ec}{3 \pi r \eta_{\rm g}} \left( \frac{u_{\rm sh}}{c} \right) B_1 t_{\rm age} \nonumber \\
	 \approx 1.72 \times 10^{14}~{\rm eV} \left( \frac{u_{\rm sh}}{0.01c} \right) \left( \frac{B_1}{1 {\rm \mu G}} \right) \left( \frac{t_{\rm age}}{200 {\rm yr}} \right)~~.
\label{eq:emax}
\end{eqnarray}
The time evolution of the shock velocity of a supernova remnant in a uniform medium is approximately given by 
\begin{eqnarray}
	u_{\rm sh} \propto 
	\left\{
	\begin{array}{l}
		(t/t_{\rm ST})^0 ~~(t \le t_{\rm ST}) \\
	 	(t/t_{\rm ST})^{-3/5} ~~(t \ge t_{\rm ST})
	\end{array}
	\right.
\end{eqnarray}
where $t_{\rm ST} \approx 200~{\rm yr}$ is the transition time when the evolution of supernova remnants changes from the free expansion phase to the adiabatic expansion phase(Sedov-Taylor phase) \citep{taylor50,sedov59,mckee95}.
Then, the time evolution of the maximum energy is 
\begin{eqnarray}
	E_{\rm max,age,1} \propto 
	\left\{
	\begin{array}{l}
		(t/t_{\rm ST})^1 ~~(t \le t_{\rm ST}) \\
	 	(t/t_{\rm ST})^{2/5} ~~(t \ge t_{\rm ST})
	\end{array}
	\right.~~.
\end{eqnarray}
Interestingly, the maximum energy increases with time even for $t\ge t_{\rm ST}$ where the shock velocity decreases with time. 
This feature cannot be seen in DSA at parallel shocks.
However, for realistic supernova remnants, we cannot expect a significant increase in the maximum energy after the free expansion phase ($t\ge t_{\rm ST}$) 
because the time dependence, $E_{\rm max,age,1} \propto t^{2/5}$, is weak and 
the acceleration time is not decided by the upstream residence time 
after the shock velocity becomes slower than $u_{\rm sh,c}\sim 0.01c$ (see  equation (\ref{eq:ushc})). 
When the acceleration time is determined by the downstream residence time, the maximum energy decreases with time \citep{ptuskin03,ohira12}.
As a result, the maximum energy that SNRs can accelerate in their lifetime is roughly estimated by equation~(\ref{eq:emax}). 
To obtain the maximum energy of PeV scale,  our model requires the upstream magnetic field strength of  at least $10~\mu$G. 
It should be noted that the maximum energy could be limited by escape processes from the acceleration region. 
In addition, the escape processes are important to understand the source spectrum of Galactic CRs \citep[e.g.][]{ptuskin05,ohira10}.  
How particles accelerated at the perpendicular shock escape from the acceleration region has never been studied. 
We are going to investigate this interesting problem in the next paper.

\subsection{Maximum energy limited by synchrotron cooling}
The maximum energy of relativistic electrons accelerated by the DSA is often limited by the synchrotron cooling.
The energy loss by the cooling during each residence time is $\dot{E}_{\rm syn,1(2)}\Delta t_{1(2)}$, where $\dot{E}_{\rm syn,1(2)}=4 e^4 E^2 B_{1(2)}^2/(9 m_{\rm e}^4 c^7)$ is the energy loss rate of synchrotron radiation in each {\rm region}.
The condition that the synchrotron cooling during the upstream residence time determines the maximum energy is $\dot{E}_{\rm syn,1}\Delta t_{1}>\dot{E}_{\rm syn,2}\Delta t_{2}$ which is reduced to
\begin{eqnarray}
\frac{u_{\rm sh}}{c}  &>& \frac{4 r}{3\pi}  \left( \frac{B_2}{B_1} \right) \\ \nonumber
	             &\approx& 1.70 \left(\frac{B_2}{B_1}\right)~~.
\end{eqnarray}
This inequality is never fulfilled because $B_2/B_1>1$ and $u_{\rm sh}/c<1$.
Under the condition of this study, therefore, the maximum energy of electrons is determined by the synchrotron cooling in the downstream region.
This result does not depend on which residence time determines the acceleration time.
In other words, information about the upstream region cannot be obtained from the maximum energy of electrons and the cutoff frequency of the synchrotron radiation as long as the maximum energy is limited by the synchrotron cooling. 
On the other hand, as shown in Figure \ref{fig:spectrum2_comparison}, if the maximum energy is limited by a finite acceleration time, the cutoff shape of the energy spectrum depends on whether the acceleration time is decided by the upstream residence time or not. 
Therefore, by analyzing the cutoff structure in detail \citep[e.g.][]{yamazaki14,yamazaki15}, 
we could investigate whether particles are accelerated at the perpendicular shock or not.

\subsection{Effects of magnetic field fluctuations in the upstream region}
In this study, we use only the uniform magnetic field in the upstream region and the plane shock surface.
In reality, however, there is not only the uniform magnetic field but also the magnetic field fluctuation and the shock surface has a finite curvature.
The acceleration time can be longer than that of our model because of the magnetic field fluctuation in the upstream region.
In this study, all of the shock surface becomes the superluminal shock region.
However, the shock surface is divided into two regions, which are the superluminal shock region and the subluminal shock region.
In the subluminal shock, particles can move along the magnetic field line in the upstream region.
Therefore, we can expect that the residence time in the upstream region becomes longer than the gyro period and that the acceleration time also becomes longer than that of our model.
Furthermore, the subluminal shock region would be important for the escape process.
We are preparing the paper about this topic and are going to submit it soon.

\subsection{A non-Bohm type Diffusion Coefficient}
So far, we have assumed that the downstream diffusion coefficient $D_2$ is proportional to the particle momentum to represent the Bohm-like diffusion. 
Here, we discuss a more general case, $D_2\propto p^a$ where $a\neq 1$. 
As long as the upstream residence time is longer than the downstream residence time, the acceleration time is given by the first term of Equation (\ref{eq:Tacc_our_model}), and the cutoff shape has the sharp structure as shown in the red histogram of Figure \ref{fig:spectrum2_comparison}. 
Thus, in this case, any results do not depend on the downstream diffusion coefficient.
However, for $a>1$, the downstream residence time eventually larger than the upstream one as particles are accelerated. 
Then, the cutoff shape changes from the sharp structure to the broad structure as the maximum momentum becomes large. 
Once the downstream residence time becomes longer than the upstream one, the acceleration time is given by 
$(3r^2/(r-1)) D_2/u_{\rm sh}^2$ and 
the broad cutoff shape depends on the diffusion coefficient \citep{yamazaki15}. 

\section{Summary} 
\label{sec:summary}
We have studied the acceleration time and the energy spectrum of particles accelerated at the perpendicular shock.
We considered the condition that the magnetic fluctuation in the upstream and the downstream regions is weak and sufficiently strong, respectively.
Under this condition, the motion of particles in our model is the gyration in the upstream region and Bohm diffusion in the downstream region. 
We derived the theoretical acceleration time in our model.
We showed the dependence of the shock velocity for the upstream residence time.
For the uniform magnetic field in the upstream region, our theoretical acceleration time is in good agreement with the simulation results. 
In addition, the energy spectrum of particles accelerated at the perpendicular shock is the same as the standard DSA prediction. 
We simultaneously realized the rapid acceleration and the canonical spectrum, $dN/dp\propto p^{-2}$, even if there is no magnetic amplification in the upstream region.
In terms of the energy spectrum, we revealed that the spectral shape of the cutoff changes whether the upstream residence time in the upstream region is longer than the downstream residence time or not. 

\acknowledgments
We thank M. Hoshino, T. Amano and S. J. Tanaka for valuable comments. 
Numerical computations were carried out on Cray XC50 at Center for Computational Astrophysics, National Astronomical Observatory of Japan. 
S.K. is supported by Yoshida Scholarship Foundation. 
Y.O. is supported by MEXT/JSPS Leading Initiative for Excellent Young Researchers. 
This work is supported by JSPS KAKENHI Grant Number JP16K17702 (Y. O.), JP19H01893(Y. O.) and JP18H01232(R. Y.). 
R.Y. deeply appreciate Aoyama Gakuin University Research Institute for helping our research by the fund.




\end{document}